\documentclass[preprint,12pt,3p]{elsarticle}

\makeatletter
\def\ps@pprintTitle{%
 \let\@oddhead\@empty
 \let\@evenhead\@empty
 \def\@oddfoot{\centerline{\thepage}}%
 \let\@evenfoot\@oddfoot}
\makeatother

\usepackage[doublespacing]{setspace}
\usepackage{amssymb}
\usepackage{graphicx}
\usepackage[space]{grffile}
\usepackage{latexsym}
\usepackage{amsfonts,amsmath,amssymb}
\usepackage{url}
\usepackage[utf8]{inputenc}
\usepackage{hyperref}
\hypersetup{colorlinks=false,pdfborder={0 0 0}}
\usepackage{textcomp}
\usepackage{longtable}
\usepackage{multirow,booktabs}

\begin{document}

\begin{frontmatter}

\title{Inter-layer and Intra-layer Heat Transfer in Bilayer/Monolayer Graphene van der Waals Heterostructure: Is There a Kapitza Resistance Analogous?}

\author[1-1,1-2]{Ali Rajabpour}
\address[1-1]{Advanced Simulation and Computing Laboratory, Mechanical Engineering Department, Imam Khomeini International University, Qazvin, Iran.}
\address[1-2]{School of Nano-Science, Institute for Research in Fundamental Sciences (IPM), Tehran, Iran.}

\author[2]{Zheyong Fan}
\address[2]{COMP Centre of Excellence, Department of Applied Physics, Aalto University, Helsinki, Finland.}

\author[3]{S. Mehdi Vaez Allaei\footnote[1]{E-mail address: smvaez@ut.ac.ir}}
\address[3]{Department of Physics, University of Tehran,Tehran, Iran.}
\renewcommand{\thefootnote}{\fnsymbol{footnote}}

\begin{abstract}
Van der Waals heterostructures have exhibited interesting physical properties. In this paper, heat transfer in hybrid coplanar bilayer/monolayer (BL-ML) graphene, as a model layered van der Waals heterostructure, was studied using non-equilibrium molecular dynamics (MD) simulations. Temperature profile and inter- and intra-layer heat fluxes of the BL-ML graphene indicated that, there is no fully developed thermal equilibrium between layers and the drop in average temperature profile at the step-like BL-ML interface is not attributable to the effect of Kapitza resistance. By increasing the length of the system up to 1 $\mu$m in the studied MD simulations, the thermally non-equilibrium region was reduced to a small area near the step-like interface. All MD results were compared to a continuum model and a good match was observed between the two approaches. Our results provide a useful understanding of heat transfer in nano- and micro-scale layered 2D materials and van der Waals heterostructures. 
\end{abstract}

\begin{keyword}
Van der Waals heterostructure \sep Bilayer/monolayer graphene \sep Kapitza resistance \sep Molecular dynamics  \sep Heat transfer	
\end{keyword}

\end{frontmatter}

\section{Introduction}

Promising electrical, mechanical and thermal characteristics of graphene and other layered materials will make them appropriate candidates for electronics, optomechanics and thermal devices in the near future. Besides single-layer graphene which has found a wide ranges of applications, its combination with other 2D and/or 3D materials with heterogeneous structures has found many other applications, making possible the manufacturing of nanoscale devices\cite{Geim2013}. Depending on the structural characteristic of a graphene heterostructure, e.g. a step-like bilayer/monolayer graphene interface, compared to covalent bonds, van der Waals interactions can lead to special properties which are considerably different from those of pristine monolayer and even bilayer graphene sheets in some cases\cite{MonoBiMonoYin2013,Rameshti2015,Berahman2014,Rajabpour2012}. 

In experimentally produced bilayer graphene or graphene flake, it is often observed that the top-layer is interrupted suddenly while the bottom-layer still keeps going\cite{Tian_2013,Huang_2008}. Based on promising electronic properties of hybrid bilayer/monolayer graphene experimentally synthesized by epitaxial growth, investigating thermal properties of these structures is important in thermal management of electronic devices as well as thermoelectric applications
\cite{Nakanishi_2010,Koshino_2010,Tian_2013,NatMatt2011BiMono,MonoBilayerGrapheneSiC2012,PhysRevXMonoBilayerGraphene2014}. In such a hybrid layered nanostructure, the step-like configurations can be considered as an interface or intercept which imposes significant effects on electrical properties. But from the phononic heat transfer point of view, the difference in properties between two sides of the interface may lead to major/minor influences on heat carriers \cite{TTM}. It also brings all features of heat transfer at an interface, including phonon scattering and Kapitza resistance\cite{POLLACK_1969, Swartz_1989}. One side of the interface consists of just covalent bonds, with the other side having van der Waals interactions as well. This combination, especially when both sides of the system are connected to different heat baths, may suggest a drop in the temperature profiles of the system at the interface, as it is widely known as Kapitza resistance \cite{Rajabpour_2010,Rajabpour_2011,Gordiz_2011,Rajabpour_2014,Chalopin_2014,Liu_2014}. But in bilayer/monolayer hybrids (see Figure 1) where each layer can be weakly/strongly coupled to another layer, the step-like interface may not lead to a Kapitza resistance in different types of configurations. The combination of intra-layer and inter-layer heat transfers makes the transfer phenomenon more complicated than that in non-layered media. For understanding the underlying physics, one should make clear what is going on at the interface and measure the amount of heat passing through within and between  layers across the entire system.

With thermal resistance at the interface being well-studied, the present research investigates, as the main research question, heat transfer in a bilayer/monolayer step-like graphene interface in a van der Waals heterostructure, but the results may be generalized to any other layered media with heterogeneous interactions. This combination makes the problem more interesting from fundamental and technological points of view. Due to the presence of strong intra-layer covalent bonds and weak inter-layer Lennard-Jones interaction, one should consider the effect of system size on inter-layer/intra-layer thermal transport. By increasing the system size and accumulation of intra-layer interactions, cross-plane mode may become more important in bilayer region, leading to interesting crossovers in heat transfer; this is more evidently observed when one moves from a small system to a macroscopic one. This could be addressed by atomistic simulations; however, in a very large system, we should compare the results with those obtained from continuum models. Thus, comparing continuum models with-large scale atomistic simulations is important.

In this paper, temperature distribution and heat flux variation are evaluated in a non-equilibrium thermal system consisting of a step-like bilayer/monolayer graphene interface. In spite of the similarities in preliminary results with Kapitza resistance, details of temperature profile and energy flux variation reveal no Kapitza resistance. The idea was supported by investigating the heat transfer: intra-layer (in-plane), inter-layer (cross-plane), and also through-interface heat transfers. A continuum model is also provided and its results are compared with MD results. To make valid comparisons, MD simulations are carried out in different length scales ranging from 0.1 $\mu m$ to 1 $\mu$m. The rest of this paper has been organized as follows: In Sections 2, details of MD simulation and setup of heat transfer in BL-ML graphene are introduced. Section 3 begins with presenting a discussion on the MD results in terms of temperature profile and heat flux distribution in BL-ML graphene and proceeds to continuum modelling of heat transfer in BL-ML graphene, ending up presenting the results of continuum model.

\section{MD simulation details}
Atomistic structure of hybrid bilayer/monolayer graphene is schematically shown in Figure 1. The bilayer graphene is in AB stacking mode which is more stable than AA stacking mode. As it is shown in Figure 1, the bilayer graphene was interrupted at its half-length from where it was extended as a monolayer graphene. Employing freezing atoms at two ends of the system, boundary conditions were fixed along x-direction. Length of the structure was varied between 0.1 $\mu$m and 1 $\mu$m and periodic boundary conditions were applied in $y$-direction. The width was set to 5 nm, which was large enough to avoid finite size effects in $y$-direction. Inter-layer thickness in bilayer graphene was initially assumed to be 3.4 $\mathring{A}$. The Tersoff potential \cite{Tersoff1989} re-parametrized for graphene by Lindsay and Broido \cite{Lindsay2010} and the Lennard-Jones potential with parameters $\epsilon = 2.39$ meV and $\sigma= 0.34$ nm \cite{Girifalco_2000} were used to to describe the intra-layer and the inter-layer interactions, respectively. It has been recently shown that, Tersoff potential is suitable for studying phonon transport in graphene as it properly predicts thermal conductivity of graphene, as compared to corresponding experimental data\cite{Mortazavi_2012,Xu2014}. Regarding these potential functions, Newton's second law of motion was integrated using velocity Verlet algorithm with a time step of 1 fs. In order to investigate heat transfer in the BL-ML graphene, the system was firstly relaxed at 300 K for 1 ns using a Nos\'{e}-Hoover thermostat chain \cite{Nose_1984,Hoover_1985,Tuckerman}, which is a well-known and accurate method for temperature control. Using this thermostat, hot and cold baths were established at the two ends of the structure at temperatures of $T_h$ and $T_c$, respectively. After 5 ns of being in steady-state condition, data collection was started for calculating temperature distribution across the system. For this purpose, the structure was divided to several computational slices and the temperature profile ($T$) across each slice was calculated using $T=\frac{2}{3k_BN}\sum_{i=1}^N \frac{1}{2}m_iv_i^2$, where $N$ is the number of atoms in each slice, $k_B$ is the Boltzmann constant, $m_i$ is the atomic mass and $v_i$ is velocity of each atom within the slice. Final temperature of each slice was calculated by time-averaging over 5 ns. All of the MD simulations were performed using the GPUMD (Graphics Processing Units Molecular Dynamics) code \cite{GPUMD}. 

\begin{figure}[ht]
\includegraphics[angle=0,scale=0.65]{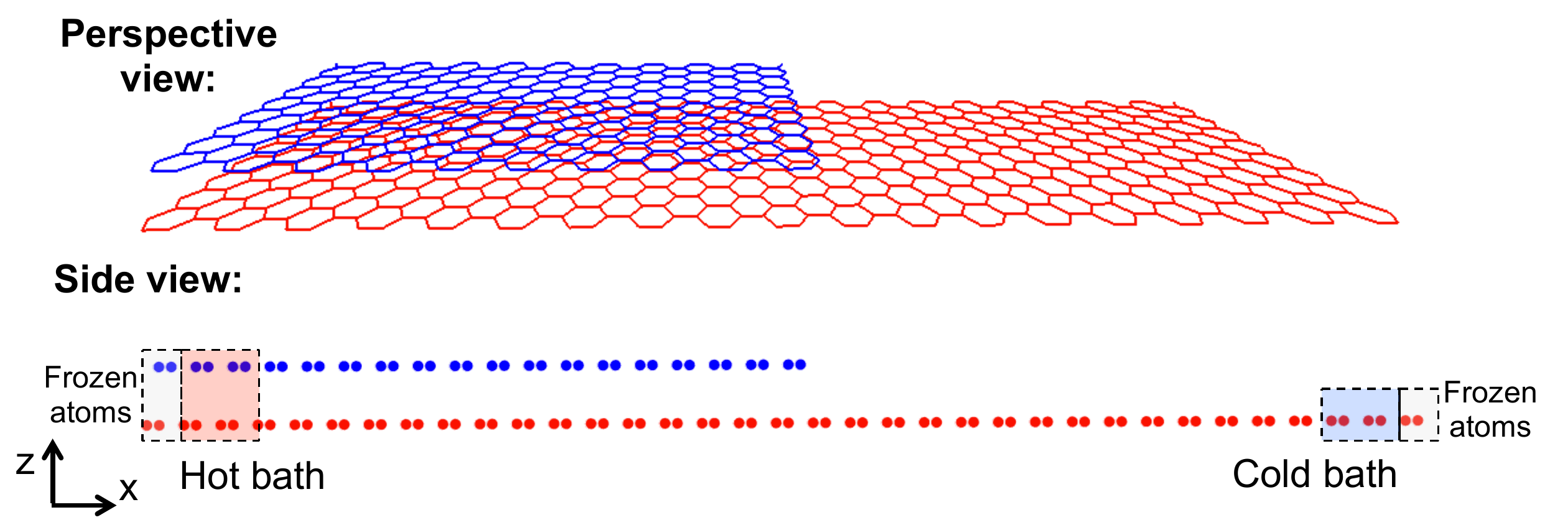}
\caption{Schematic view of atomistic structure of hybrid bilayer/monolayer graphene. The graphene half-layer and full-layer are marked in blue and red, respectively. Non-equilibrium heat transfer is generated by applying hot and cold baths at two ends of the structure at temperatures of $T_h$ and $T_c$, respectively.
 }
\end{figure}

\section{Results and discussion}

\subsection{Kapitza resistance at interface}
In non-equilibrium molecular dynamics method, each computational slice is assumed to be locally in equilibrium. Considering small size of a slice and long phonon mean free path, this assumption is challenging; however, the method has been widely used for predicting thermal conductivity of materials\cite{Schelling_2002, Wang_2009, Bagri_2011, Zhong_2011,Wei_2011, Mortazavi_2015}. In the studied hybrid bilayer/monolayer graphene in the present research, two slicing methods were considered (methods (i) and (ii) in Figure 2). In method (i), the bilayer graphene was considered as an integrated structure with thermal equilibrium between its layers. In method (ii), however, no thermal equilibrium was assumed between the layers, i.e. each layer had its own temperature profile.

In order to investigate the accuracy of each slicing method, temperature profiles were calculated in a structure with the length of $L$ = 0.1 $\mu$m, $T_h$ = 320 K and $T_c$ = 280 K. The results of slicing methods (i) and (ii) are shown in Figure 2(a) and (b), respectively. With method (i), there is a temperature drop at intercept which may be attributed to a Kapitza resistance between the two parts of the structure upon which no temperature difference is seen between the layers. In method (ii), temperature differences between layers of the bilayer part reveals that, thermal equilibrium between the layers has not been reached. The reason for such differences will be discussed in the next section. Thus, the slicing method (i) does not lead to correct results as it returns unrealistic Kapitza resistance at intercept.
 
\begin{figure}[ht]
\begin{center}
\includegraphics[angle=0,scale=1]{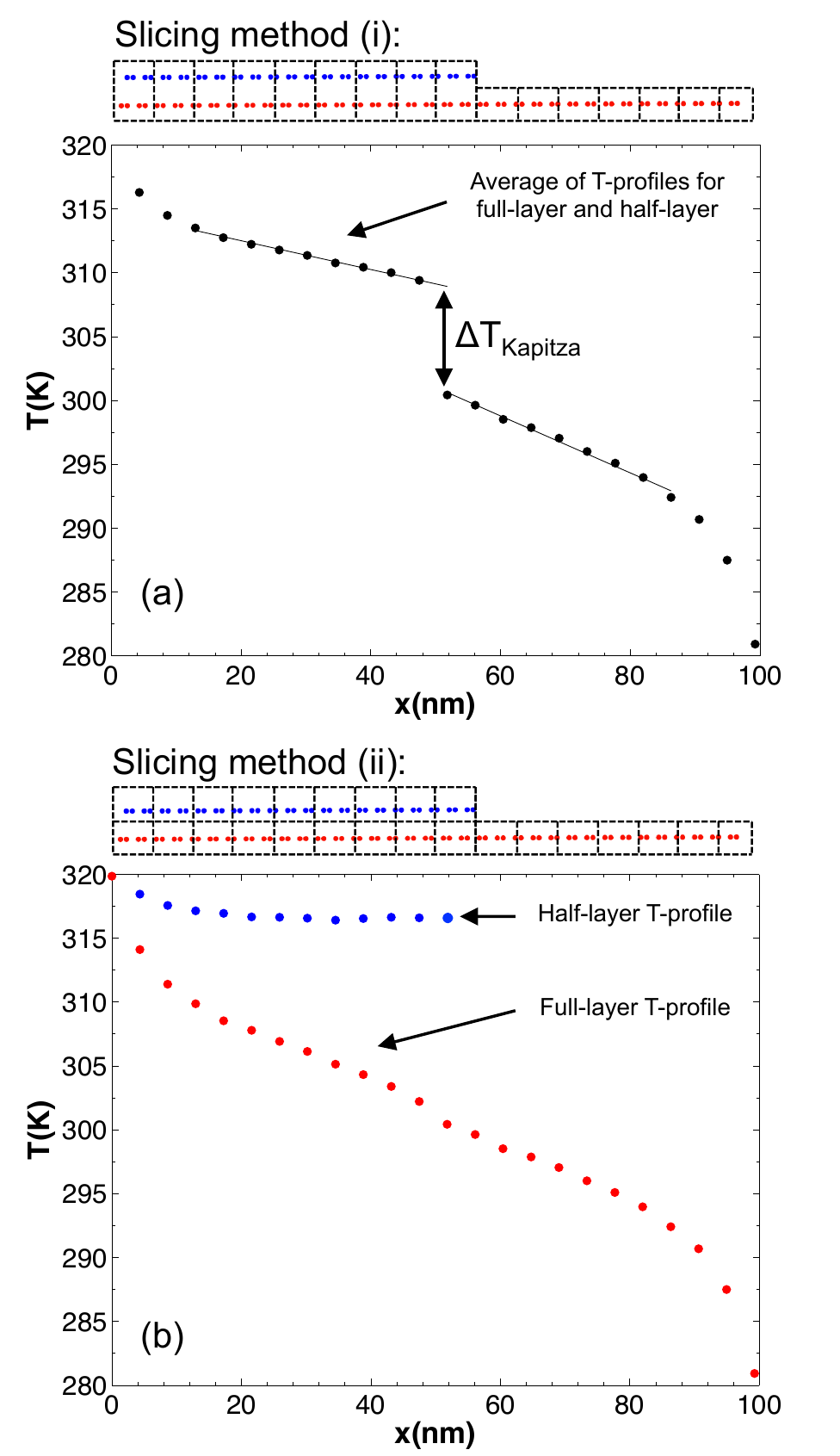}
\caption{Temperature profiles in the BL-ML graphene extracted from two slicing methods (i) and (ii). The method (i) leads to appearance of a non-realistic 
Kapitza resistance, where, half-layer graphene is disconnected as it is shown in (a) while the method (ii) reveals that there is no thermal equilibrium 
between half-layer and full-layer as depicted in (b).  
 }
\end{center}
\end{figure}

\subsection{MD simulation results}
Temperature profiles for a hybrid bilayer/monolayer graphene with dimensions of $L$ = 0.5 and 1 $\mu$m are shown in Figure 3. As mentioned in the previous section, the layers of the bilayer part differed in temperature, especially for smaller length of the system. This was due to low thermal conductivity between layers of the bilayer graphene (cross-plane thermal conductivity) described by weak van der Waals atomic interactions compared to high in-plane thermal conductivity of graphene layers which could be addressed by strong intra-layer covalent interactions \cite{Ni_FLG_APL}. Since the top layer of the bilayer graphene was only connected to the hot bath, the heat current must go through the layers of the bilayer graphene with low inter-layer thermal conductivity, developing some temperature differences between different graphene layers. 
In order to better understand the heat transfer in the BL-ML graphene, heat flux distribution across the structure was calculated. For this purpose, inter-layer and intra-layer heat fluxes are shown in Figure 4. Intra-layer heat flux from a computational slice (e.g. A) to another slice (e.g. B) in each layer was calculated using microscopic definition of heat flux as follows\cite{Fan_decompose}:

\begin{equation}
\begin{split}
q_{A \rightarrow B}=-\frac{1}{2} \sum_{i \in A} \sum_{j \in B} \left<  \left( \frac{\partial \sum_{k\neq i} U_{ik}}{\partial \vec{r}_{ij} } \cdot \vec{v}_j -   \frac{\partial \sum_{k\neq j} U_{jk}}{\partial {\vec{r}_{ji}}}  \cdot \vec{v}_i \right) \right>
\end{split}
\end{equation}
where $\vec{r}_{ij}\equiv \vec{r}_j-\vec{r}_i$ is the difference in position between atom $i$ and atom $j$, $\vec{v}$ is atomic velocity, and $U_{ik}$ is the bond energy between atoms $i$ and $k$. Inter-layer heat flux could be calculated from the difference in intra-layer heat flux between two adjacent slices according to the energy conservation law. As shown in Figure 4, the inter-layer heat flux between two layers of the bilayer graphene increases with length; this is due to decreased heat flux along the top layer (half-layer). So, the heat flux through layers reaches its maximum value at the end of the half-layer graphene, where the maximum temperature difference between the top and bottom layers takes place. The heat flux in the bottom graphene layer (full-layer) also increases along the length, reaching a maximum value at $x=L/2$ and remaining constant for $x>L/2$.

\begin{figure}[t]
\begin{center}
\includegraphics[angle=0,scale=1]{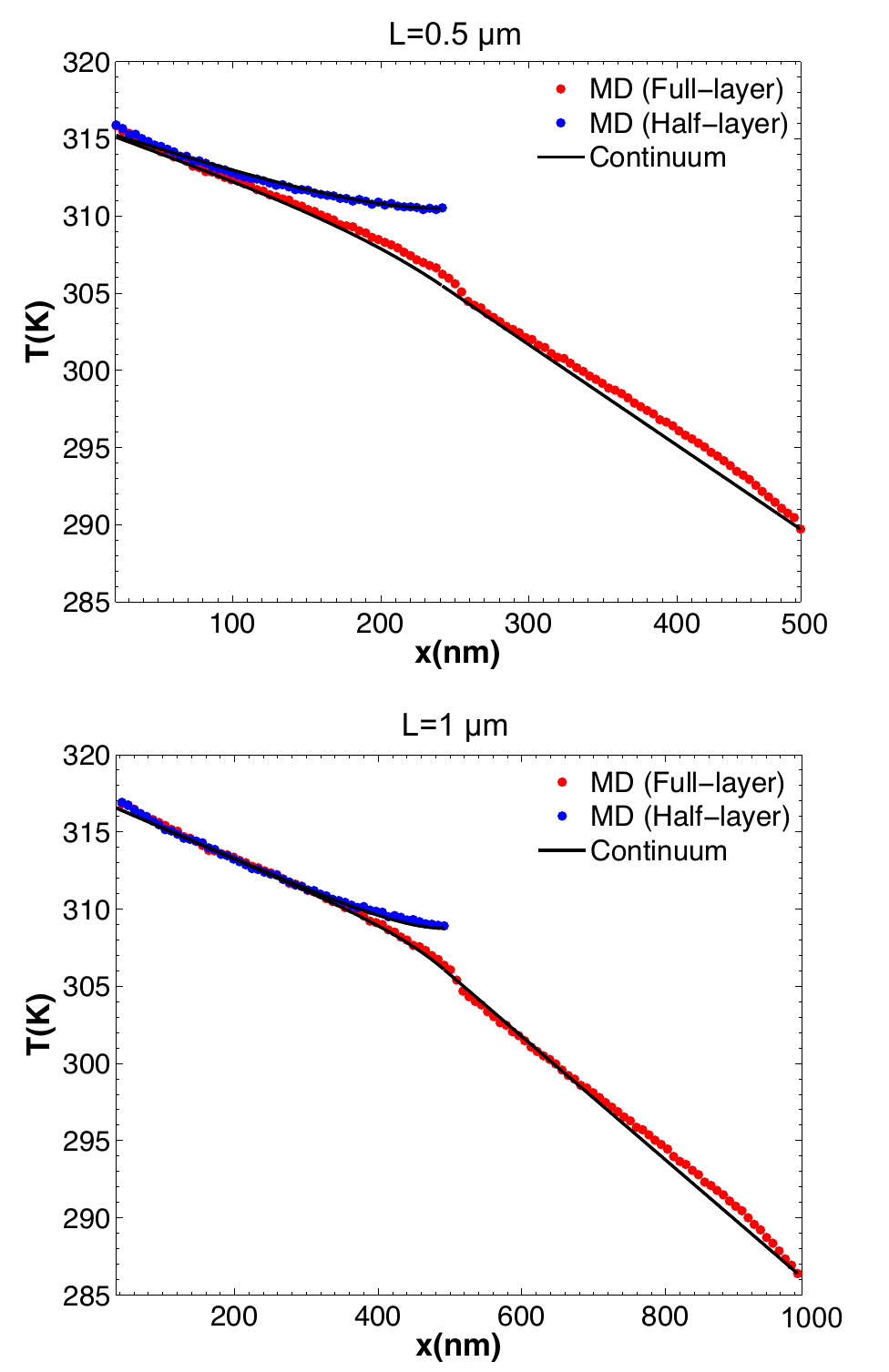}
\caption{Temperature profile in the BL-ML graphene for $L=0.5$ and $1\mu m$. Discrete points refer to MD results and continuous black lines correspond to continuum results.}
\end{center}
\end{figure}

\begin{figure}[t]
\begin{center}
\includegraphics[angle=0,scale=0.8]{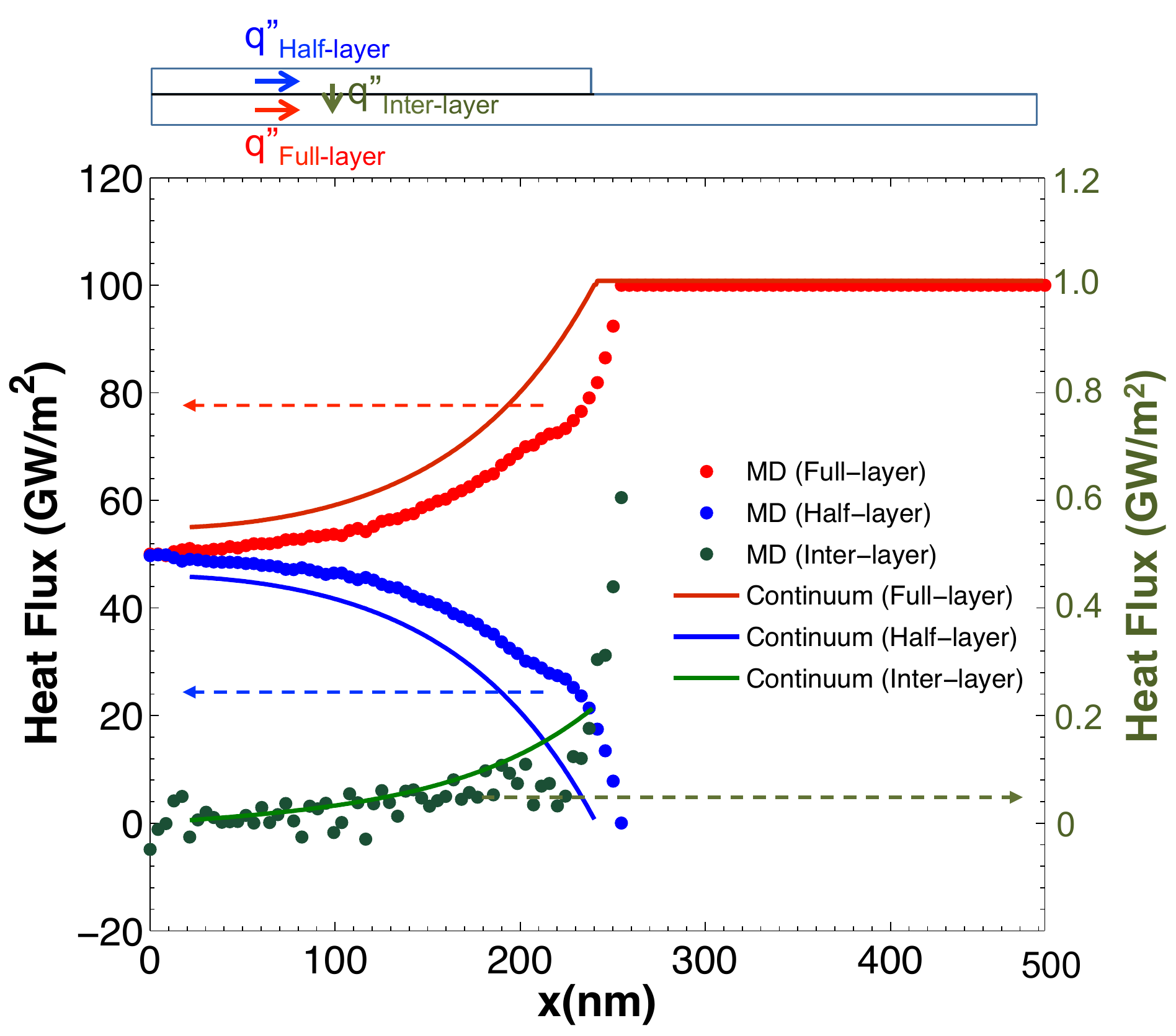}
\caption{Intra-layer and inter-layer heat flux distributions calculated via MD simulation (points) and continuum modelling (lines) for the BL-ML graphene with L=0.5 $\mu$m. The heat flux in the full-layer increases with $x$, reaching to its maximum value at $x=L/2$, while it follows a decreasing trend with increasing $x$ in the half-layer. Based on the energy conservation law, the difference of $q"_{full-layer}$ and $q"_{half-layer}$ at each $x$ equals to $q"_{inter-layer}$.}
\end{center}
\end{figure}

\subsection{Continuum results}
In order to see what happens when system length is very large, a continuum model was considered for solving heat conduction equation in BL-ML graphene structure. The layers were considered to be continuous with in-plane thermal conductivity of $\kappa_{ip}$. In order to model cross-plane thermal conductivity, a thin layer with the thermal conductivity of $\kappa_{cp}$ was considered in between the plates. The $\kappa_{ip}$ and $\kappa_{cp}$ were extracted from MD simulations based on the linear portion of temperature gradient in the monolayer graphene and temperature differences between layers of the bilayer graphene, respectively, and considered as inputs into the continuum model. As a length-dependent parameter, $\kappa_{ip}$ was calculated to be 1546 W/mK and 1950 W/mK for L=0.5 $\mu$m and 1 $\mu$m, respectively, which are in good agreements with previous reports on thermal conductivity of graphene \cite{Mortazavi_2012,Xu2014}. The value of $\kappa_{cp}$ was found to be 0.015 W/mK, i.e. clearly much lower than $\kappa_{ip}$. Thickness of each layer was set to $t$ = 3.4 $\mathring{A}$. In this way, by solving the heat conduction equation across the structure, temperature distribution across the half-layer ($T_{hl}$) and the full-layer ($T_{fl}$) were obtained as follows:

\begin{equation}
\begin{split}
\frac{d^2T_{hl}(x)}{dx^2}-P(T_{hl}-T_{fl}(x)) &=0, \qquad 0\leq x \leq L/2 \\
\frac{d^2T_{fl}(x)}{dx^2}+P(T_{hl}-T_{fl}(x))&=0, \qquad 0\leq x \leq L   
\end{split}
\end{equation}
with boundary conditions as: \\
\begin{equation}
\begin{split}
At \ x&=0:  \ \quad T_{hl}=T_{fl}=T_H, \\
At \ x&=L/2: \quad \frac{dT_{hl}}{dx}=0, \ \frac{dT_{fl}}{dx} \ \text{and} \ T_{fl} \ \text{are} \ \text{continuous functions.} \\   
At \ x&=L:  \ \quad T_{fl}=T_C, 
\end{split}
\end{equation}
where $T_H$ and $T_C$ are hot and cold bath temperatures, respectively, and $P=\frac{\kappa_{cp}}{\kappa_{ip}t^2}$. The two above-mentioned equations are coupled, so that those must be solved simultaneously. By algebraic summation of equations and analytically solving them, temperature profiles across the layers were obtained as follows:

\begin{equation}
\begin{split}
T_{hl}(x)&=T_H-A\left(\frac{x}{2}+\frac{\sinh(\sqrt{2P}x)} {2\sqrt{2P}\cosh(\sqrt{P/2}L)} \right), \qquad 0\leq x \leq L/2 \\
T_{fl}(x)&=T_H-A\left(\frac{x}{2}-\frac{\sinh(\sqrt{2P}x)} {2\sqrt{2P}\cosh(\sqrt{P/2}L)} \right), \qquad 0\leq x \leq L   \\
\end{split}
\end{equation}
where 

\begin{equation}
\begin{split}
A=\left( \frac{T_H-T_C}{L/2} \right) / \left(\frac{3}{2}+\frac{\tanh(\sqrt{P/2}L)}{L\sqrt{2P}} \right)
\end{split}
\end{equation}

The temperature profiles obtained from the analytical solution of energy equations are shown in Figure 3 (continuous black lines). Accordingly, a good agreement was found between MD results and continuum model results. The temperature jump at the intercept could be calculated from the continuum model in terms of the structure length and applied temperature difference to the thermal baths, as follows:

\begin{equation}
\begin{split}
\Delta T \vert_{x=L/2} =(T_{hl}-T_{fl})\vert_{x=L/2}=\frac{ 2(T_{hl}-T_{fl}) \tanh(\sqrt{P/2}L) } {  3L\sqrt{P/2}+\tanh(\sqrt{P/2}L)  }
\end{split}
\end{equation}

The heat flux calculated from the continuum simulation is also shown in Figure 4 (continuous lines). According to the continuum results, the same trend to MD results was followed by heat flux in the BL-ML graphene. 
 
\section{Summary}
In summary, temperature distribution and inter-layer and intra-layer heat fluxes were calculated in a hybrid bilayer/monolayer graphene system at non-equilibrium steady state using molecular dynamics simulations and continuum heat conduction modelling. Accordingly, it was revealed that, in systems of shorter length, a temperature drop is seen between the layers, not only at the step-like interface, but also in a considerable part of the bilayer graphene. Thus, defining a Kapitza resistance at the planar intercept is an unrealistic practice. It was also shown that, the temperature drop will gradually vanish when the system increases in size. All MD results were compared to the continuum model and a good match was observed between the two approaches. Our findings can provide useful understanding concerning heat transfer not only in bilayer/monolayer graphene, but also in other layered 2D materials and van der Waals heterostructures.

\section*{Acknowledgement}
We acknowledge the computational resources provided by Imam Khomeini International University, IPM, Aalto Science-IT project and Finland's IT Center for Science (CSC).

\bibliographystyle{model1-num-names}

\end{document}